\documentclass[conference]{IEEEtran}
\usepackage{graphicx}
\usepackage{float}
\usepackage{amsmath}
\usepackage{amssymb}
\usepackage{commath} % norm and absolute value
\usepackage{algorithm}
\usepackage{algpseudocode}
\usepackage{pifont}
\usepackage{pdfpages}
\usepackage{graphicx}
\usepackage{float}
\usepackage{color}
\usepackage{subcaption}
\usepackage{parskip}

\ifCLASSINFOpdf
\else
\fi

\begin{document}
\title{Optimal Design of Virtual Inertia and Damping Coefficients for Virtual Synchronous Machines}

\author{\IEEEauthorblockN{Atinuke Ademola-Idowu, \ Baosen Zhang}
\IEEEauthorblockA{Electrical Engineering Department,\\
University of Washington, Seattle, WA 98195}
Email: \{aidowu, zhangbao\}@uw.edu}

\maketitle

\begin{abstract}
Increased penetration of inverter-connected renewable energy sources (RES) in the power system has resulted in a decrease in available rotational inertia which serves as an immediate response to frequency deviation due to disturbances. The concept of virtual inertia has been proposed to combat this decrease by enabling the inverters to produce active power in response to a frequency deviation like a synchronous generator.
In this paper, we present an algorithm to optimally design the inertia and damping coefficient required for an inverter-based virtual synchronous machine (VSM) to participate efficiently in the inertia response portion of primary frequency control. We design the objective function to explicitly trade-off between competing objectives such as the damping rate the the frequency nadir. Specifically, we formulate the design problem as a constrained and regularized H2 norm minimization problem, and develop an efficient gradient algorithm for this non-convex problem. This proposed algorithm is applied to a test case to demonstrate its performance against existing methods.
\end{abstract}

% \fi
\IEEEpeerreviewmaketitle

\section{Introduction}
In a traditional electric power system network with only conventional generators, the rotating synchronous machines connected to the network possesses kinetic energy as a result of their rotating mass and releases it as an immediate response in the event of a power imbalance to reduce the rate of frequency decline \cite{machowski2011power}. As the electric power grid transitions from this traditional state to a mix of conventional generators and inverter-connected RES, this immediate response capability by the synchronous machine is reduced. This results in an increased rate of change of frequency (ROCOF) and consequently, a higher frequency deviation which results in a low frequency nadir, that is, the maximum frequency deviation \cite{tielens2012grid}. The ROCOF, frequency nadir and settling time/frequency are important frequency response metrics in the power systems network.

To combat this problem, various techniques that utilize inverter-connected RES and energy storage systems (ESS) have been proposed. One of such techniques is called De-loading. This provides a reserve margin in the wind turbine or Photo-voltaic (PV) by operating on a reduced power level as compared to its maximum power extraction point \cite{dreidy2017inertia}. Other techniques include designing the inverter to behave like a synchronous generator as in the case of a synchroconverter \cite{tamrakar2017virtual,zhong2011synchronverters}, using a swing equation based approach that computes the swing equation every control cycle to emulate inertia \cite{sakimoto2011stabilization,alipoor2015power} or emulating the inertia by monitoring the ROCOF and frequency deviation, and producing power proportional to that change as in the case of the VSM\cite{tamrakar2017virtual}. In the later case, for a variable speed wind turbine such as a doubly fed induction generator, a suitable controller can be used to release the kinetic energy stored in the rotating blades based on the ROCOF and frequency deviation \cite{diaz2014participation,dreidy2017inertia}. A similar controller can be used for an ESS, the difference will be in the source of the energy.

The components of interest in the design the controller for the VSM is the gain blocks mimicking the inertia and damping coefficients values. There is a need for this values to be optimally selected depending on the current operating point and the units committed so as to efficiently participate in the inertia response. A variant of this problem was considered in in \cite{borsche2017placement} and \cite{poolla2017optimal} where the focus was instead on the placement of the virtual inertia though the same reasoning applies to the problem being considered. In \cite{borsche2017placement} the optimal virtual inertia and damping placement was computed with explicit time-domain constraints on the frequency response metrics. while in \cite{poolla2017optimal}, only the optimal virtual inertia placement was considered with the performance metrics in this case being the $\mathcal{H}_2$ norm as a measure of network coherency. The shortcomings of the first approach is its approximations and computational burden while for the second approach, the $\mathcal{H}_2$ norm does not provide a means of trading-off between competing system objectives. We improve on this shortcoming by augmenting the $\mathcal{H}_2$ norm objective with a tuning parameter that gives control over the frequency performance metrics and can be adjusted to determine the inertia and damping coefficient values for any frequency performance requirements we choose to satisfy.

The contributions of this paper are as follows, we provide a systematic way of optimally designing the virtual inertia and damping coefficient of a two loop VSM that emulates the inertia response of a generator based on the ROCOF and frequency deviation. We derive a tuning parameter to augment the $\mathcal{H}_2$ norm problem thereby giving us control over how much of the requirements we want to satisfy as the $\mathcal{H}_2$ norm performance metric cannot by itself simultaneously satisfy the essential time domain frequency performance requirements: slow ROCOF, low frequency nadir and fast settling time. We also propose an explicit computationally efficient gradient for the non-convex optimization problem. The proposed method is applied to a test case for validation and the simulation result in time domain performed as expected.

The remaining section are organized as follows: Section \ref{Modeling} summarizes the modeling of the electric power systems and the VSM. Section \ref{Performance} describes the frequency response requirements in the power systems and objective functions that can quantify these requirements. Section \ref{formulation} presents the optimization problem and the proposed gradient for the computation. Section \ref{results} shows the application of the algorithm to a test case.

\section{Modeling} \label{Modeling}
\subsection{Power System Model}
In a power systems network, the electromechanical dynamics of a synchronous generator is governed by the swing equation \cite{machowski2011power}:
\begin{equation} \label{swing}
\begin{aligned}
& \frac{d \delta_i}{dt} = \bigtriangleup \omega_i\\
& M_i \frac{d \bigtriangleup\omega_i}{dt} + D_i \frac{d \delta_i}{dt} = P_{m,i} - P_{e,i}, \quad \forall i \in \{1,  \cdots, n\}.
\end{aligned}
\end{equation}
where $\delta_i$ is the rotor angle, $\omega_i$ is the rotor speed, $M_i$ is the generator inertia constant, $D_i$ is the damping coefficient, $P_m$ is the mechanical input power to the generator and $P_e$ is the electric power output from the generator and $n$ is the number of generators.

The swing equation in (\ref{swing}) can be linearized around an operating point to obtain a model suitable for analyzing the response of the generators when subjected to disturbances. This model can be represented in first order state space form as in (\ref{swing_matrix}):
\begin{equation} \label{swing_matrix}
 \begin{bmatrix}
  \dot{\bigtriangleup \delta} \\
  \dot{\bigtriangleup \omega}
 \end{bmatrix}
  =
\underbrace{
 \begin{bmatrix}
  0 & I \\
  -M^{-1}L & -M^{-1}D
 \end{bmatrix}
 }_{ = A}
 \begin{bmatrix}
  \bigtriangleup \delta \\
  \bigtriangleup \omega
 \end{bmatrix}
 +
\underbrace{
 \begin{bmatrix}
  0\\
  M^{-1}
  \end{bmatrix}
  }_{ = B}
  \bigtriangleup P
\end{equation}
where M = diag($M_i$) and D = diag($D_i$). $\bigtriangleup P$ represents an imbalance between the mechanical and electrical power. Under assumptions for DC power flow, $\bigtriangleup P_i$ can be simplified to:
\begin{equation}
\begin{aligned}\label{DCPF}
\bigtriangleup P_i = \sum_{j \neq n} L_{ij}(\bigtriangleup \delta_i - \bigtriangleup \delta_r) = \sum_{j \neq n} L_{ij} \bigtriangleup \delta_{ir}
\end{aligned}
\end{equation}
where $\bigtriangleup \delta_r$ is the reference angle which is the angle of designated swing generator. For ease of notation, we drop the subscript $r$ and use $\bigtriangleup \delta_i$ to represent the relative angle. Under the assumptions made, $L$ is the network susceptance matrix (shunt admittance ignored) given by $L_{ii} = \sum_{i\neq j} b_{ij}$ and $L_{ij} = -b_{ij}$.

% In the small signal model in eqn (\ref{swing_matrix}), the network has been reduced to an equivalent network using Kron's reduction where the all nodes have generators and are connected to one another.

\subsection{Virtual Synchronous Machines (VSM)}
\begin{figure}[!htbp]
  \centering
  \includegraphics[width = \linewidth]{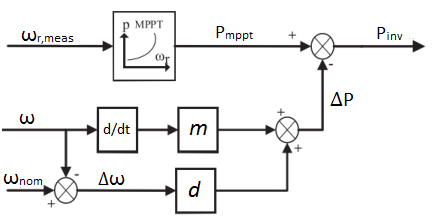}
  \caption{A two control loop VSM utilizing the ROCOF and frequency deviation measurement to produce active power for inertia emulation \cite{dreidy2017inertia}.}
  \label{fig:Emul}
\end{figure}

% \subsection{Inverter Modeling / Inertia Emulation}
As discussed earlier, the main idea behind a VSM, is to emulate the inertia response by controlling the inverter to respond to changes in frequency \cite{tamrakar2017virtual}. The change in output power produced by the inverter in response to the frequency change functions as described in Fig. \ref{fig:Emul} and is governed by $\bigtriangleup P = \tilde{\mathbf{m}} \frac{d \bigtriangleup \omega}{dt} + \tilde{\mathbf{d}} \bigtriangleup \omega$. A review of this technique can be found in \cite{morren2006wind}\cite{sakimoto2011stabilization}\cite{van2009grid}.
% \todo{equation need to be part of a sentence}
% \begin{equation}
% \begin{aligned}\label{PVSM}
% \bigtriangleup P = \tilde{M} \frac{d \bigtriangleup \omega}{dt} + \tilde{D} \bigtriangleup \omega
% \end{aligned}
% \end{equation}

The values of gain block $\tilde{m}$ and $\tilde{d}$ are the virtual inertia and damping coefficient to be designed. The choice of these values determine how efficiently the inverter responds and should therefore be optimally selected to achieve the frequency response performance metrics specified by the operators.

% This two loop control is better than the single loop control solely based on the ROCOF to avoid .... The advantage of this form of inertia provision is that the VSM becomes dispatchable and responds to changes in system frequency \cite{tamrakar2017virtual}

\section{Performance Metrics} \label{Performance}
% \todo{It would be good to explicitly state the performance metrics, rather than hiding them in the discussion. For example, have }

% \noindent \todo{\textbf{$H_2$ norm:} This measures ...}

% \noindent \todo{\textbf{Something about $M_i$} This measures ...}

When an event occurs in a power system network, frequency stability is maintained through the primary, secondary and tertiary frequency controls. The automatic inertia response which releases the rotational kinetic energy in the machines and the proposed virtual inertia response falls under the primary control and occurs in the first few seconds ($\sim$10s) of an event \cite{eto2011use}. In a power systems network, the major frequency performance metrics evaluated to determine the sufficiency of the available frequency controls are the ROCOF, frequency nadir and settling time/frequency \cite{eto2011use}. It is desired to have a slow ROCOF, low frequency nadir and fast settling time/frequency. To quantify these, we consider the rotating kinetic energy produced by the generators during an event, this can be represented as the quadratic function $z = \frac{1}{2} \bigtriangleup \dot{\omega}^T M \bigtriangleup \dot{\omega}$ and in matrix form as in equation (\ref{output_matrix}):

\begin{equation}\label{output_matrix}
\resizebox{.3\hsize}{!}{$z =
\underbrace{
  \begin{bmatrix}
    0_n & M^{\frac{1}{2}}
    \end{bmatrix}
}_{ = C}
\begin{bmatrix}
  \bigtriangleup \delta \\
  \bigtriangleup \omega
 \end{bmatrix}$}
\end{equation}

We want to minimize this energy released in the event of a disturbance while ensuring that the system is stable. This can be used as a performance metric when trying to emulate this behavior in a inverter-based VSM. A minimized energy requirement connotes a reduced area under the frequency curve. A suitable and well developed control performance metrics that captures this objective is the $\mathcal{H}_2$ norm (commonly interpreted as the impulse-to-energy gain)\cite{poolla2017optimal} \cite{tegling2015price}. Depending on the system structure, the system objectives might be competing as in our case. For example, consider a second order mechanical system (mass-spring-damper system), for a given damping value, a lower mass results in a higher overshoot but faster settling time while a higher mass results in a smaller overshoot but a slower settling time.

To provide a means of trading-off between these competing system objectives or explicitly controlling the frequency performance requirements, consider the swing equation in (\ref{swing}), which can be re-written to obtain the ROCOF (and by extension, the maximum frequency deviation) as:
\begin{equation}\label{ROCOF}
\frac{d \bigtriangleup \omega_i}{dt} = \frac{1}{M_i}(\bigtriangleup P_i - D_i\bigtriangleup \omega_i).
\end{equation}
If our goal is to minimize the ROCOF ($\frac{d \bigtriangleup \omega_i}{dt}$) in (\ref{ROCOF}), then we maximize $M_i$ such that it is larger than ($\bigtriangleup P_i - D_i\bigtriangleup \omega_i$). Conversely, if in a certain scenario a faster settling time is required (which implies a higher frequency), then as explained earlier for a second order system where $M_i \propto \frac{1}{\omega_i^2}$, $M_i$ has to be minimized in order to maximize $\omega_i$. Since these two requirements depend on $m$ in opposite ways, the tuning control is therefore set as $\beta \norm{\mathbf{m}}^2_2$ where the sign of $\beta$ varies depending on which of the objectives we intend to satisfy.

\section{Problem Formulation} \label{formulation}
\subsection{Optimization Problem}
The total objective function will be a combination of the two objective described in section \ref{Performance}. The first is the $\mathcal{H}_2$ norm which seeks to strike a balance between two competing objectives: minimizing frequency nadir and ROCOF, and minimizing the settling time, while the second represented by $\beta \norm{\mathbf{m}}^2_2$ gives preference on the control over these competing objective. The design variables $\mathbf{m}$ and $\mathbf{d}$ are components of the matrix $A, B$ and $C$ in (\ref{swing_matrix}) and (\ref{output_matrix}).
The $\mathcal{H}_2$ norm of a linear system with impulse response $G(t) = Ce^{At}B + D$ is computed using \cite{zhou1996robust}:

% \begin{subequations}
% \begin{align} \label{H2_SS}
% &  \norm{G}_{\mathcal{H}_2}^2 = \mathrm{Tr}\left[\int_{0}^{+\infty} G(t)^T G(t) dt \right] = \mathrm{Tr}\left[B^T Q B \right]\\
% & \norm{G}_{\mathcal{H}_2}^2 =  \mathrm{Tr}\left[\int_{0}^{+\infty} G(t) G(t)^T dt \right]= \mathrm{Tr}\left[C P C^T \right]
% \end{align}
% \end{subequations}
\begin{subequations}
\begin{align} \label{grada}
& J = \norm{G}_2^2 = \mathrm{Tr} (C P C^T)\\
\label{gradc}
& J = \norm{G}_2^2 = \mathrm{Tr} (B^T Q B)
\end{align}
\end{subequations}
where $P$ and $Q$ are the observability and controllability gramian and satisfy the Lyapunov equation in (\ref{gradb}) and its dual in (\ref{gradd}) respectively.

\begin{subequations}
\begin{align} \label{gradb}
& AP + PA^T +BB^T = 0 \\
\label{gradd}
& A^T Q + QA + C^T C = 0
\end{align}
\end{subequations}

The optimization problem is then formulated as:
\begin{subequations} \label{Opt_prob}
\begin{align}
\label{opt_obj}
& \underset{\mathbf{m},\mathbf{d}}{\text{minimize}}
& & J_T(\mathbf{m},\mathbf{d}) = \norm{G(\mathbf{m},\mathbf{d})}^2_2 + \beta \norm{\mathbf{m}}^2_2\\
& \text{subject to}
\label{opt_m}
& & \underline{\mathbf{m}} \leq \mathbf{m} \leq \overline{\mathbf{m}}\\
\label{opt_d}
& & & \underline{\mathbf{d}} \leq \mathbf{d} \leq \overline{\mathbf{d}}\\
\label{opt_P}
& & & \resizebox{.65\hsize}{!}{$A(\mathbf{m},\mathbf{d})P + PA(\mathbf{m},\mathbf{d})^T +B(\mathbf{m})B(\mathbf{m})^T = 0$}\\
\label{opt_Q}
& & & \resizebox{.65\hsize}{!}{$A^T(\mathbf{m},\mathbf{d})Q + QA(\mathbf{m},\mathbf{d})^T +C(\mathbf{m})^TC(\mathbf{m}) = 0$}\\
& & & \resizebox{.25\hsize}{!}{$P \succ 0 ; \quad Q \succ 0$}
\end{align}
\end{subequations}
The upper and lower bound limit on $\mathbf{m}$ and $\mathbf{d}$ in (\ref{opt_m}) and (\ref{opt_d}) is determined by the RES source or ESS considered. For wind RES, this is determined by the wind turbine rotor speed which is proportional to the wind speed, and for a battery ESS, it is determined by the capacity and state of charge. These value typically vary depending on the time of the day and operating conditions but we assume that they are fixed for the optimization period which is in seconds.

The constraints on the Lyapunov equation and its dual in (\ref{opt_P}) and (\ref{opt_Q}) makes the problem non-convex and difficult to solve using existing algorithms. This non-convexity can be observed in the multiplication of matrix $A$ and $P$, and $A^T$ and $Q$ which consists of the design and unknown variables. Despite the non-convexity of the objective function, it is smooth \cite{van2008h2} and therefore an explicit gradient can be derived and the optimal values of the coefficient obtained using (\ref{T2}).

\begin{equation}
\begin{aligned}\label{T2}
\mathbf{\alpha}^{k+1} = \mathrm{Proj}_{\mathcal{C}}[\alpha^k - \gamma\bigtriangledown J(\alpha^k)]
\end{aligned}
\end{equation}
where $\mathbf{\alpha} = [\mathbf{m} \; \mathbf{d}]^T $ and $\mathcal{C}$ is the set of feasible $m$ and $d$ values. A projected gradient descent technique can be used for the optimization since the constraints on $m$ and $d$ form a boxed constraint.

\subsection{Gradient Computation}
Inspired by \cite{wilson1970optimum}, a computationally efficient gradient of the objective function in (\ref{Opt_prob}) is obtained as follows:
Let $\alpha$ be any variable in matrix $A, B$ or $C$. Taking the derivative of (\ref{grada}) w.r.t $\alpha$, we have:
\begin{equation}
\begin{aligned}\label{PJ1}
\frac{\partial J}{\partial \alpha} = \mathrm{Tr} \left(\frac{\partial P}{\partial \alpha} (C^T C) \right) + \mathrm{Tr} \left(P \frac{\partial (C^T C)}{\partial \alpha}\right)
\end{aligned}
\end{equation}
Re-arranging (\ref{gradd}) and substituting into (\ref{PJ1}):
\begin{equation*}
\begin{aligned}\label{T1}
\frac{\partial J}{\partial \alpha} = \mathrm{Tr} \left(\frac{\partial P}{\partial \alpha} -(A^TQ + QA) \right) + \mathrm{Tr} \left(P \frac{\partial (C^T C)}{\partial \alpha}\right)
\end{aligned}
\end{equation*}

\begin{equation}
\begin{aligned}\label{PJ2}
\frac{\partial J}{\partial \alpha} = -2\mathrm{Tr} \left(\frac{\partial P}{\partial \alpha} QA \right) + \mathrm{Tr} \left(P \frac{\partial (C^T C)}{\partial \alpha}\right)
\end{aligned}
\end{equation}

Taking the derivative of (\ref{gradb}) w.r.t $\alpha$, post multiplying by $Q$ and taking the trace of the resulting equation:
\begin{equation}
\begin{aligned}\label{Der2}
-2 \mathrm{Tr} \left(\frac{\partial P}{\partial \alpha} QA \right) = 2 \mathrm{Tr} \left(\frac{\partial A}{\partial \alpha} PQ \right) + \mathrm{Tr} \left(\frac{\partial (BB^T)}{\partial \alpha} Q\right)
\end{aligned}
\end{equation}
Finally, substituting (\ref{Der2}) into (\ref{PJ2}) gives the gradient:
\begin{equation}
\begin{aligned}\label{PJ3}
\frac{\partial J}{\partial \alpha} = 2 \mathrm{Tr} \left(\frac{\partial A}{\partial \alpha} PQ \right) + \mathrm{Tr} \left(\frac{\partial (BB^T)}{\partial \alpha} Q\right) + \mathrm{Tr} \left(P \frac{\partial (C^T C)}{\partial \alpha}\right)
\end{aligned}
\end{equation}
The gradient of $\beta \norm{\mathbf{m}}^2_2$ is given as $2 \beta \mathbf{m}$.
% Therefore:
% \begin{equation}
% \begin{aligned}\label{PJ4}
% \bigtriangledown J(\alpha^k) = 2 \mathrm{Tr} \left(\frac{\partial A}{\partial \alpha} PQ \right) + \mathrm{Tr} \left(\frac{\partial (BB^T)}{\partial \alpha} Q\right) + \mathrm{Tr} \left(P \frac{\partial (C^T C)}{\partial \alpha}\right) + 2 \beta m
% \end{aligned}
% \end{equation}
% when $\alpha = m$ and

\begin{figure*}[!htbp]
\centering
    %\captionsetup[subfigure]{aboveskip=-8pt} % spaceing between image and caption for top row
    \begin{subfigure}[h]{0.45\textwidth}
    \centering
    \hspace*{-0.95cm}
    \includegraphics[width = \linewidth,keepaspectratio]{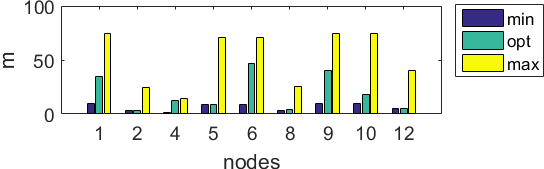}
    \caption{} % comment this out if you don't want (a), (b) etc
    \label{fig:Result_a5a}
  \end{subfigure}
  \begin{subfigure}[h]{0.45\textwidth}
    \centering
    \hspace*{-0.9cm}
    \includegraphics[width = \linewidth,keepaspectratio]{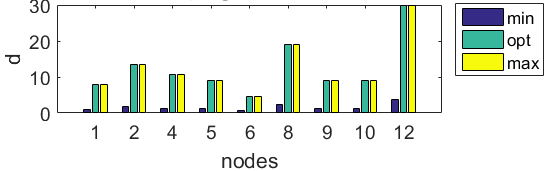}
    \caption{} % comment this out if you don't want (a), (b) etc
    \label{fig:Result_a5b}
  \end{subfigure}
  \begin{subfigure}[h]{0.45\textwidth}
    \centering
    \hspace*{-0.95cm}
    \includegraphics[width = 3.3in,keepaspectratio]{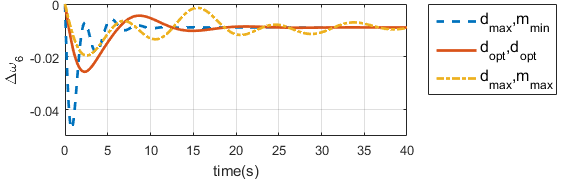}
    \caption{} % comment this out if you don't want (a), (b) etc
    \label{fig:Result_a5c}
  \end{subfigure}
  \begin{subfigure}[h]{0.5\linewidth}
    \centering
    % \hspace*{-0.9cm}
    \includegraphics[width = 3in]{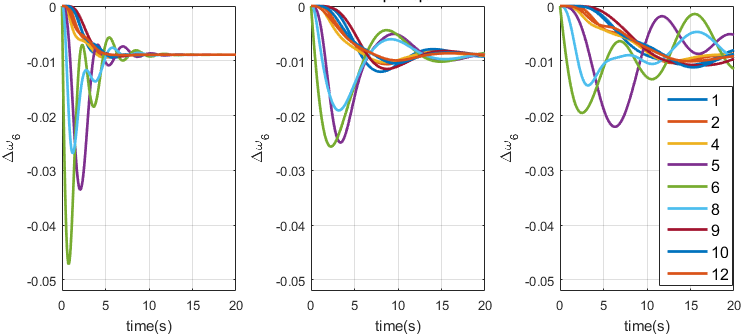}
    \caption{} % comment this out if you don't want (a), (b) etc
    \label{fig:Result_a5d}
  \end{subfigure}
\caption{This assumes the location of disturbance is unknown. (a) Optimal inertia coefficient distribution across all nodes, (b) Optimal damping coefficient distribution across all nodes. (c) Frequency deviation at node 6 due to step input. The optimal response balances between a fast and slow ROCOF to give smoother response (d) Frequency deviation at all nodes due to step input. \textit{LHS}: combination of $d_{max}m_{min}$, \textit{MID}: combination of $d_{opt}m_{opt},[\textit{RHS}$]: combination of $d_{max}m_{max}$. It shows how the optimal coefficients results in a balanced smoother response in all the nodes.}
\label{fig:equal_likely_disturbance}
\end{figure*}

\section{Results} \label{results}
\begin{figure}[!htbp]
\centering
\hspace*{-0.9cm}
\includegraphics[width = 3.2in]{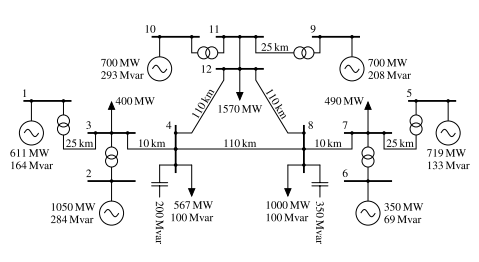}
\caption{One line diagram of a 12 bus three-area test case \cite{poolla2017optimal}.}
\label{fig:12Bus}
\end{figure}

The optimization problem in (\ref{Opt_prob}) is implemented using a modified 12 bus three-area test case \cite{poolla2017optimal}\cite{kundur1994power} shown in Fig. \ref{fig:12Bus}. The network is reduced to an equivalent network by removing the static load buses (3, 7 and 11) using the Krons reduction method. Also, the available inertia and load damping at the remaining buses are reduced to model a scenario of high penetration of RES. We assume that there are inverter-connected RES at each bus which implies that in the matrix representation of the swing equation in (\ref{swing_matrix}), $M = \hat{\mathbf{m}} + \mathbf{m}$ and $D = \hat{\mathbf{d}} + \mathbf{d}$ where $\hat{\mathbf{m}}$ and $\hat{\mathbf{d}}$ are the coefficients of the synchronous generator and load damping and $\mathbf{m}$ and $\mathbf{d}$ are the variables of the inertia control loop in Fig. \ref{fig:Emul} to be designed.

% In the following analysis, the optimal result is compared with assigning a combination of maximum damping and minimum inertia ($d_{max}m_{min}$), and maximum damping and inertia ($d_{max}m_{max}$). These combinations were chosen because $d_{max}m_{min}$ gives a faster response time but bigger overshoot, while $d_{max}m_{max}$ gives a slower response time but smaller overshoot.

In the following analysis, the optimal result is compared with assigning the maximum damping and minimum inertia ($d_{max}m_{min}$) which gives a faster response time but bigger overshoot, and maximum damping and inertia ($d_{max}m_{max}$) which gives a slower response time but smaller overshoot.

A step input is applied to one of the nodes, in this case, node 6. Fig. \ref{fig:Result_a5a} and \ref{fig:Result_a5b} show the optimal inertia and damping coefficient distribution across the nodes while Fig. \ref{fig:Result_a5c} and Fig. \ref{fig:Result_a5d} show the step response at the disturbed node 6 and at all the nodes respectively. From these step responses, it can be seen that the optimal inertia and damping coefficient values result in a time response that achieves a balance between having a fast settling time but high frequency nadir and fast ROCOF, and a slow settling time but low frequency nadir and slow ROCOF. In this situation where it is not possible to simultaneously achieve these objectives, a Pareto front is achieved by the optimization problem in (\ref{Opt_prob}).

\begin{figure}[!htbp]
  \centering
  % \hspace*{-0.95cm}
  \includegraphics[width = 3.2in]{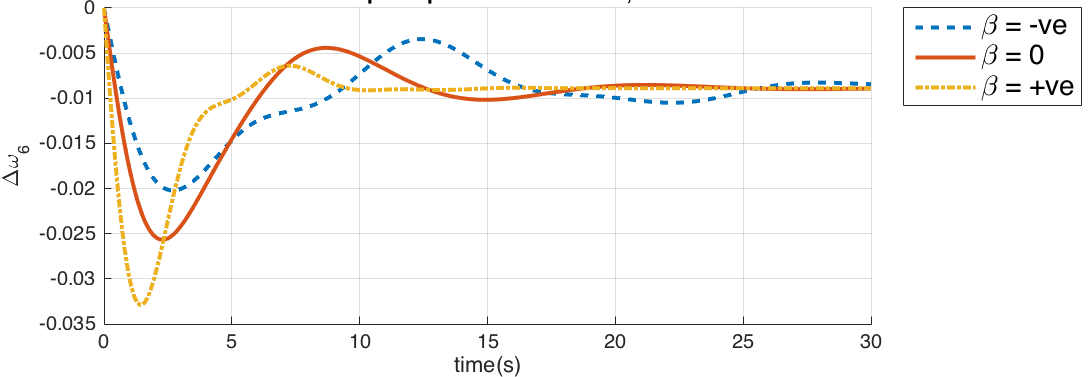}
  \caption{Frequency deviation at node 6 for different values of $\beta$ due to step input. A negative $\beta$ value results in a lower ROCOF (which is preferred) and smaller frequency deviation but slower time response while a positive $\beta$ value results in a higher ROCOF and larger frequency deviation but faster time response.}
  \label{fig:Result_diff2}
\end{figure}

The effect of $\beta$ in (\ref{Opt_prob}) can be observed in Fig. (\ref{fig:Result_diff2}). Varying $\beta$ determines the extent of the trade-off between the ROCOF/frequency nadir and settling time, in the case of providing a fast inertia response, a low ROCOF is desired. The $\beta$ therefore controls the amount inertia in the system to achieve the desired the ROCOF.
% It can also be seen in fig. (\ref{fig:Result_a5c}) and fig. (\ref{fig:Result_a5d}) that the resulting optimal response is smooth. The total inertia and damping coefficients assigned to each of the inverters are $\sum m_i = 173.54$ and $\sum d_i = 112.50$ respectively.

\begin{figure*}[!htbp]
\centering
    %\captionsetup[subfigure]{aboveskip=-8pt} % spaceing between image and caption for top row
    \begin{subfigure}[h]{0.45\textwidth}
    \centering
    \hspace*{-0.95cm}
    \includegraphics[width = \linewidth,keepaspectratio]{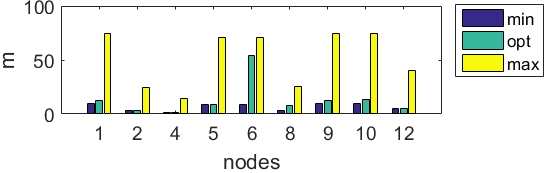}
    \caption{} % comment this out if you don't want (a), (b) etc
    \label{fig:Result_d5a}
  \end{subfigure}
  \begin{subfigure}[h]{0.45\textwidth}
    \centering
    \hspace*{-0.9cm}
    \includegraphics[width = \linewidth,keepaspectratio]{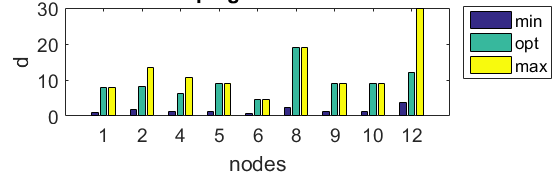}
    \caption{} % comment this out if you don't want (a), (b) etc
    \label{fig:Result_d5b}
  \end{subfigure}
  \begin{subfigure}[h]{0.45\textwidth}
    \centering
    \hspace*{-0.95cm}
    \includegraphics[width = 3.5in,keepaspectratio]{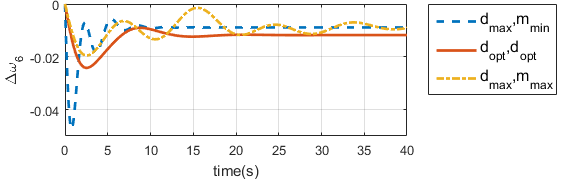}
    \caption{} % comment this out if you don't want (a), (b) etc
    \label{fig:Result_d5c}
  \end{subfigure}
  \begin{subfigure}[h]{0.5\linewidth}
    \centering
    % \hspace*{-0.9cm}
    \includegraphics[width = 3in]{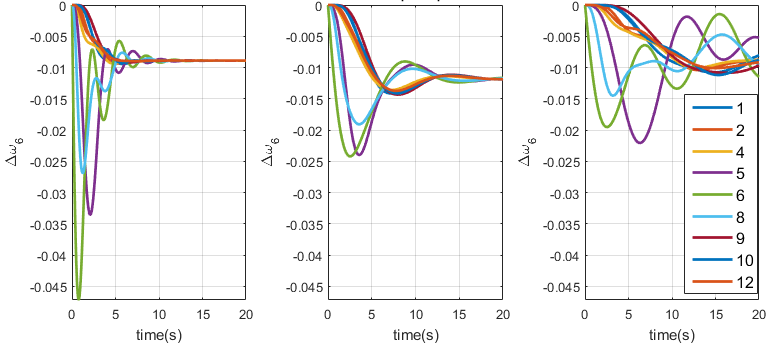}
    \caption{} % comment this out if you don't want (a), (b) etc
    \label{fig:Result_d5d}
  \end{subfigure}
\caption{This assumes the location of disturbance is known. (a) Optimal inertia coefficient distribution across all nodes, (b) Optimal damping coefficient distribution across all nodes. (c) Frequency deviation at node 6 due to step input. The optimal response still achieves a balanced response despite reduction in inertia and damping values (d) Frequency deviation at all nodes due to step input. \textit{LHS}: combination of $d_{max}m_{min}$, \textit{MID}: combination of $d_{opt}m_{opt},[\textit{RHS}$]: combination of $d_{max}m_{max}$. It shows how the optimal coefficients results in a balanced smoother response in all the nodes.}
\label{fig:known_disturbance}
\end{figure*}

We consider next, how the inertia and damping coefficient distribution across the nodes changes if the location of the disturbance is known beforehand. For this case, we assume the disturbance is likely to occur at node 6, we therefore have that $\bigtriangleup P$ in (\ref{swing_matrix}) is given by $\bigtriangleup P = \eta\Delta(t)$. The $\Delta(t)$ represents an impulse input and $\eta$ is represented by the standard basis $\mathbf{e_i}$ in the direction of node 6 and is pre-multiplied by the matrix $B$ in (\ref{swing_matrix}) before the optimization. The results are shown in Fig. \ref{fig:known_disturbance} and it can be seen that there is a reduction in the optimal inertia and damping coefficient distribution across the nodes for a similar disturbance and time response compared to Fig. \ref{fig:equal_likely_disturbance} where no knowledge of the disturbance is known. The optimal inertia value is highest for the disturbed node while the optimal damping value is highest for not only the disturbed, but also the surrounding nodes.

Based on these analysis, it can be seen that the control weight $\beta$ gives the ability to control the frequency performance requirements. Also, computing an optimal value for the $\tilde{\mathbf{m}}$ and $\tilde{\mathbf{d}}$ requires some knowledge of the system state and history. If this is unknown, the optimal values are allocated in a robust way to cater for all disturbance scenarios. The $\tilde{\mathbf{m}}$ and $\tilde{\mathbf{d}}$ design variables can be recomputed and the new values assigned to the gain blocks of the inverter controller at every operating point. This will be required as the operating point and parameters of the power systems network changes depending on the units committed.

\section{Conclusion}
We considered the problem of optimally designing the virtual inertia and damping coefficient of a two loop VSM that emulates the inertia response of a generator based on the ROCOF and frequency deviation. We used an augmented $\mathcal{H}_2$ norm objective function to incorporate the time domain frequency response requirements and provide an explicit computationally efficient gradient to compute the coefficient values. We applied this method to a test case and our time domain simulation performed as expected from the theory. We also showed via simulation, how the knowledge of a potential disturbance location could help in efficiently computing the coefficients.

Going forward, we would like to extend the model used to a more detailed model to ensure a more robust parameter design and also consider dynamically computing the coefficients depending on the current system condition.

\bibliographystyle{IEEEtran}
\bibliography{ref_PES}

\end{document}